\documentclass[11pt,fleqn]{article}
\usepackage{amsfonts,amssymb,latexsym,cite}
\usepackage{epsf,graphicx}

\def\noi{\noindent}

\newcommand{\Title}[1]{\noi {{\Large\bf #1}}\\[1ex]}

\newcommand{\Author}[2]{\noi{\large\bf #1}\\[2ex]\noi{\normalsize\it #2}\\}

\newcommand{\Rec}[1]{\noi {\it Received #1} \\}

\newcommand{\Abstract}[1]{\vskip 2mm \begin{center}
        \parbox{16.4cm}{\small\noi #1} \end{center}\medskip}

\newcommand{\foom}[1]{\protect\footnotemark[#1]}

\def\email#1#2{\footnotetext[#1]{e-mail: #2}\addtocounter{footnote}{1}}

\def\nq{\hspace*{-1em}}
\def\nqq{\hspace*{-2em}}
\def\nhq{\hspace*{-0.5em}}

\def\cm{\hspace*{1cm}}
\def\inch{\hspace*{1in}}

\def\ten#1{\mbox{$\times 10^{#1}$}}
\def\deg{\mbox{${}^\circ$}}                     

\def\Jl#1#2{#1 {\bf #2},\ }

\def\ApJ#1 {\Jl{Astroph. J.}{#1}}
\def\CQG#1 {\Jl{Class. Quantum Grav.}{#1}}
\def\DAN#1 {\Jl{Dokl. AN SSSR}{#1}}
\def\GC#1 {\Jl{Grav. \& Cosmol.}{#1}}
\def\GRG#1 {\Jl{Gen. Rel. Grav.}{#1}}
\def\JETF#1 {\Jl{Zh. Eksp. Teor. Fiz.}{#1}}
\def\JETP#1 {\Jl{Sov. Phys. JETP}{#1}}
\def\JHEP#1 {\Jl{JHEP}{#1}}
\def\JMP#1 {\Jl{J. Math. Phys.}{#1}}
\def\NPB#1 {\Jl{Nucl. Phys.}{B\ #1}}
\def\NP#1 {\Jl{Nucl. Phys.}{#1}}
\def\PLA#1 {\Jl{Phys. Lett.}{#1A}}
\def\PLB#1 {\Jl{Phys. Lett.}{#1B}}
\def\PRD#1 {\Jl{Phys. Rev.}{D\ #1}}
\def\PRL#1 {\Jl{Phys. Rev. Lett.}{#1}}

\def\al{&\nhq}
\def\lal{&&\nqq {}}
\def\eq{Eq.\,}
\def\eqs{Eqs.\,}
\def\beq{\begin{equation}}
\def\eeq{\end{equation}}
\def\bear{\begin{eqnarray}}
\def\bearr{\begin{eqnarray} \lal}
\def\ear{\end{eqnarray}}
\def\earn{\nonumber \end{eqnarray}}
\def\nn{\nonumber\\ {}}
\def\nnv{\nonumber\\[5pt] {}}
\def\nnn{\nonumber\\ \lal }
\def\nnnv{\nonumber\\[5pt] \lal }
\def\yy{\\[5pt] {}}
\def\yyy{\\[5pt] \lal }
\def\eql{\al =\al}

\def\dst{\displaystyle}
\def\tst{\textstyle}
\def\fracd#1#2{{\dst\frac{#1}{#2}}}
\def\fract#1#2{{\tst\frac{#1}{#2}}}
\def\Half{{\fracd{1}{2}}}
\def\half{{\fract{1}{2}}}

\def\e{{\,\rm e}}
\def\d{\partial}

\def\sign{\mathop{\rm sign}\nolimits}

\def\const{{\rm const}}
\def\eps{\varepsilon}
\def\ep{\epsilon}

\def\mn{_{\mu\nu}}
\def\MN{^{\mu\nu}}
\def\mN{_\mu^\nu}
\def\nM{_\nu^\mu}
\def\cK{{\cal K}}
\def\cV{{\cal V}}

\def\kappa{\varkappa}

\def\wt{\widetilde}
\def\tg{{\wt g}}
\def\tR{{\wt R}}

\def\oR{{\overline R}}

\def\sss{\scriptscriptstyle}

\def\mD{m_{\sss D}}

\begin{document}
\twocolumn[
\thispagestyle{empty}

\Title
   {Variations of $\alpha$ and $G$ from nonlinear multidimensional gravity}

\Author{K.A. Bronnikov\foom 1 and M.V. Skvortsova\foom 2}
       {\small Center for Gravitation and Fundamental Metrology, VNIIMS,
             46 Ozyornaya St., Moscow 119361, Russia;\\
    Institute of Gravitation and Cosmology, PFUR,
             6 Miklukho-Maklaya St., Moscow 117198, Russia}

\Rec{February 10, 2013}

\Abstract
  {To explain the recently reported large-scale spatial variations of the
  fine structure constant $\alpha$, we apply some models of
  curvature-nonlinear multidimensional gravity. Under the reasonable
  assumption of slow changes of all quantities as compared with the Planck
  scale, the original theory reduces to a multi-scalar field theory in four
  dimensions. On this basis, we consider different variants of isotropic
  cosmological models in both Einstein and Jordan conformal frames. One
  of the models turns out to be equally viable in both frames, but in the
  Jordan frame the model predicts simultaneous variations of $\alpha$ and
  the gravitational constant $G$, equal in magnitude. Large-scale small
  inhomogeneous perturbations of these models allow for explaining the
  observed spatial distribution of $\alpha$ values.
  }

] 
\email 1 {kb20@yandex.ru}
\email 2 {milenas577@mail.ru}

\section{Introduction}

  The long-standing problem of possible space-time variations of the
  fundamental physical constants (FPC) is now actively discussed on both
  theoretical and observational grounds, and in particular, in connection
  with more or less confidently observed variations of the fine-structure
  constant $\alpha$ in space and time \cite{webb1,webb2}. The first data on
  temporal changes of $\alpha$, such that $\alpha$ was in the past slightly
  smaller than now (the relative change $\delta\alpha/\alpha$ is about
  $10^{-5}$), appeared in \cite{webb1} from observations of mostly the
  Northern sky at the Keck telescope (the Hawaiian islands). In 2010, an
  analysis of new data obtained at the VLT (Very Large Telescope), located
  in Chile, and their comparison with the Keck data led to a conclusion on
  spatial variations of $\alpha$, i.e., on its dependence on the direction
  of observations. According to VLT observations in the Southern sky,
  $\alpha$ was in the past slightly larger than now. This anisotropy has a
  dipole nature \cite{webb2} and has been termed ``the Australian dipole''
  \cite{beren1}. The dipole axis is located at a declination of $-61 \pm
  9\deg$ and at a right ascension of $17.3 \pm 0,6$ hours. The deflection of
  $\alpha$ value at an arbitrary point $r$ of space from its modern value
  $\alpha_0$, measured on Earth, is, at a confidence level of $4.1 \sigma$,
\beq \label{dipole}
    \delta\alpha/\alpha_0 = (1.10 \pm 0.25) \times 10^{-6}\,
            r \cos \psi,
\eeq
  where $\psi$ is the angle between the direction of observation and the
  dipole axis, while the distance $r$ is measured in billions of light
  years \cite{webb2}.

  On the other hand, recent laboratory experiments have given the tightest
  constraints on $\alpha$ variations on Earth in the modern epoch
  \cite{rosbnd1}
\beq                                                    \label{a-clock}
    (d\alpha/dt)/\alpha = (-1.6 \pm 2.3) \times 10^{-17}\ \mbox{per year.}
\eeq
  This result is of the same order of magnitude as the tightest constraints
  obtained previously from an isotopic composition analysis of the decay
  products in the natural nuclear reactor that operated in the Oklo region
  (Gabon) about 2 billion years ago. Unlike the laboratory data, the Oklo
  results \cite{shlya} and, in particular, the tightest constraint
  \cite{fujii}
\beq                                                    \label{a-Oklo}
        d (\ln \alpha)/dt = (-0.4 \pm 0.5) \times 10^{-17}/{\rm yr}
\eeq
  rely on the assumption that during these 2 billion years the value of
  $\alpha$ changed uniformly, if changed at all. This assumption looks
  rather natural but actually follows from nowhere.

  The observed space-time distribution of $\alpha$ values is illustrated in
  Fig.\,1: on Earth, at least since the Oklo times, $\alpha$ is constant
  on the level of $\sim 10^{-17}$ per year, whereas according to the quasar
  data for about 10 billion years a variation rate can be about $10^{-15}$
  per year. Meanwhile, one cannot exclude the opportunity that the
  variations are purely spatial in nature whereas the time dependence is
  related to the finiteness of the velocity of light: being located at a
  fixed point and at fixed time, we receive signals from distant regions of
  the Universe emitted at earlier cosmological epochs, and it is therefore
  impossible to separate spatial and temporal dependences of the parameters.

\begin{figure}[h]
\centering
\includegraphics[scale=0.4]{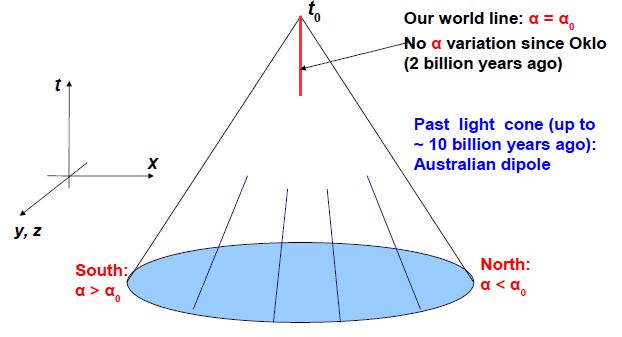}
\caption{The empirical data on variations of $\alpha$}
\end{figure}

  A number of theoretical models have been constructed in order to explain
  these variations \cite{Chiba-1, Olive-1, Olive-2, Odintsov, Barrow,
  Mariano, Mari-2}. In these approaches the variability of $\alpha$ is
  explained in the framework of general relativity with the aid of scalar
  fields whose existence, self-interaction and coupling to the electromagnetic
  field were postulated ``by hand''. In \cite{Odintsov}, it was shown that
  in $F(R)$ gravity it is possible to obtain a static effective
  (gravitational) domain wall with spatially varying $\alpha$ by postulating
  a certain nonminimal interaction between the electromagnetism and gravity.
  In \cite{we-13} it was shown that scalar fields and their interaction law
  with electromagnetism leading to variations of $\alpha$ naturally follow
  from curvature-nonlinear multidimensional gravity. It was noted that an
  advantage of multidimensional gravity in the treatment of FPC variations
  is that all such variations are explained in a unified way from spatial
  and temporal variations of the size of extra dimensions \cite{mel1, mel2}.

  In \cite{we-13}, in the approach to nonlinear multidimensional gravity
  formulated in \cite{VfromD}, a simple model was built, explaining the
  observed variations of $\alpha$. Using the methodology of \cite{VfromD},
  a particular multidimensional theory was reduced to a scalar field theory
  in 4 dimensions, which resulted in a cosmological model with accelerated
  expansion, and certain initial conditions were chosen, slightly different
  from homogeneity and isotropy. The results were obtained in the Einstein
  conformal frame, in which, by construction, the gravitational constant
  does not change.

  The present paper continues this study. We show that if we consider the
  Jordan frame as the physical (observational) one \cite{bm-predict, erice},
  then the model built in \cite{we-13} predicts a very large cosmic
  acceleration contrary to observations. We suggest another simple model
  which also fairly well describes the present accelerated stage along with
  variations of $\alpha$, but is equally viable in the Einstein and Jordan
  frames. In the latter, $\alpha$ and $G$ evolve according to the same
  law, being inversely proportional to the volume of extra dimensions.
  Unlike those of $\alpha$, variations of $G$ have not been discovered so
  far, there are only upper bounds. The tightest constraint following from
  the results of lunar laser ranging is \cite{mueller}
\beq
    {\dot G}/G = (2\pm 7) \ten{-13}\,{\rm yr}^{-1}.     \label{G-dot}
\eeq
  Since $\alpha$ variations are at most $\sim 10^{-15}$ per year, similar
  variations of $G$, if any, are in agreement with (\ref{G-dot}).

  Let us note that the methodology of \cite{VfromD}, allowing for a
  transition from a broad class of multidimensional theories of gravity with
  higher derivatives to Einstein gravity with effective scalar fields, was
  successfully applied to obtain a unified description of the early
  inflation and modern acceleration of the Universe \cite{infl}; it has
  provided a possible explanation of the origin of the Higgs field and
  solutions of some other physical and cosmological problems
  \cite{Bolokhov:2010mz, Rubin:2011jm, BRook}.

  The paper is organized as follows. Section 2 briefly describes the general
  formalism to be used. In this framework, in Section 3 we consider two
  variants of isotropic cosmological models; one of them, obtained
  previously \cite{we-13}, is viable only in the Einstein picture, while the
  other is equally viable in the Einstein and Jordan pictures. In Section 4
  we consider small large-scale inhomogeneous perturbations of these models
  and show that each of them is able to account for the observed spatial
  variations of $\alpha$. Section 5 is a brief conclusion.

\section {Basic equations}

  Consider a $(D = 4 + d_1)$-dimensional manifold with the metric
\beq                                                           \label{ds}
        ds^2 = g\mn dx^\mu dx^\nu + \e^{2\beta(x)} b_{ab} dx^a dx^b
\eeq
  where the extra-dimensional metric components $b_{ab}$ are independent of
  $x^{\mu}$, the observable four space-time coordinates.\footnote
    {Our sign conventions are:
    the metric signature $(+{}-{}-{}-)$; the
    curvature tensor
    $R^{\sigma}{}_{\mu\rho\nu} = \d_\nu\Gamma^{\sigma}_{\mu\rho}-\ldots,\
    R\mn = R^{\sigma}{}_{\mu\sigma\nu}$, so that the Ricci scalar
    $R > 0$ for de Sitter space-time and the matter-dominated
    cosmological epoch; the system of units $8\pi G = c = 1$.}

  The $D$-dimensional Riemann tensor has the nonzero components
\bear
    R\MN{}_{\rho\sigma} \eql \oR\MN{}_{\rho\sigma},       \label{Riem}
\nn
    R^{\mu a}{}_{\nu a} \eql \delta^a_b\, B\nM, \cm
    B\nM := \e^{-\beta} \nabla_\nu (\e^\beta \beta^{\mu}),        
\nn
    R^{ab}{}_{cd} \eql
      \e^{-2\beta} \oR^{ab}{}_{cd} + \delta^{ab}{}_{cd} \beta_\mu\beta^\mu,
\ear
  where capital Latin indices cover all $D$ coordinates, the bar marks
  quantities obtained from $g\mn$ and $b_{ab}$ taken separately,
  $\beta_{\mu} \equiv \d_{\mu}\beta$ and $\delta^{ab}{}_{cd}\equiv
  \delta_{c}^{a}\delta_{d}^{b}-\delta_{d}^{a}\delta_{c}^{b}$. The
  nonzero components of the Ricci tensor and the scalar curvature are
\bear
    R\mN \eql \oR\mN + d_1\, B\mN,          \label{Ric}           
\nnv
    R_a^b \eql \e^{-2\beta} \oR_a^b
                    + \delta_a^b [ \Box \beta + d_1 (\d{\beta})^2 ],
\nnv
    R \eql \oR [g] + \e^{-2\beta }\oR [b] + 2d_1 \Box \beta
\nnn \inch
           + d_1(d_1+1) (\d{\beta})^2,
\ear
  where $(\d{\beta})^2 \equiv \beta_{\mu}\beta^{\mu}$,
  $\Box = \nabla^\mu \nabla_\mu$ is the d'Alembert operator while $\oR[g]$
  and $\oR [b]$ are the Ricci scalars corresponding to $g\mn$ and $b_{ab}$,
  respectively. Let us also present, using similar notations, the
  expressions for two more curvature invariants, the Ricci tensor squared
  and the Kretschmann scalar $\cK = R^{ABCD}R_{ABCD}$
  (where capital Latin indices cover all $D$ coordinates):
\bearr
    R_{AB}R^{AB} = \oR\mn\oR\MN + 2d_1 \oR \mn B\MN + d_1^2 B\mn B\MN
\nnn\cm
        + \e^{-4\beta}\oR_{ab}\oR^{ab}
    + 2\e^{-2\beta} \oR[b] [\Box\beta + d_1 (\d{\beta})^2 ]
\nnn\cm
      + d_1 [\Box\beta + d_1 (\d{\beta})^2 ]^2,             \label{Ric2}
\yyy
    \cK = \overline{\cK}[g] + 4 d_1 B\mn B\MN + \e^{-4\beta}\overline{\cK}[b]
\nnn \quad\
      + 4 \e^{-2\beta} \oR [b] (\d{\beta})^2
         +2 d_1 (d_1-1) [(\d\beta)^2 ]^2 .                     \label{Kre}
\ear

  Suppose now that $b_{ab}$ describes a compact $d_1$-dimensional space of
  nonzero constant curvature, i.e., a sphere ($K=1$) or a compact
  $d_1$-dimensional hyperbolic space ($K = -1$) with a
  fixed curvature radius $r_0$ normalized to the $D$-dimensional
  analogue $\mD$ of the Planck mass, i.e., $r_0 = 1/\mD$ (we use the
  natural units, with the speed of light $c$ and Planck's constant $\hbar$
  equal to unity). We have
\bear
    \oR^{ab}{}_{cd} \eql K\,\mD^2\,\delta^{ab}{}_{cd},         \label{r0}
\nn
    \oR_a^b \eql K\,\mD^2\, (d_1-1) \delta_a^b,
\nn
    \oR [b] \eql K\,\mD^2\, d_1 (d_1-1) = R_b.
\ear
  The scale factor $b(x) \equiv \e^{\beta}$ in (\ref{ds}) is thus kept
  dimensionless; $R_b$ has the meaning of a characteristic curvature scale
  of the extra dimensions.

  In this geometry,
  we deal with a sufficiently general curvature-nonlinear theory of gravity
  with the action
\bear                                                         \label{act1}
     S \eql \Half \mD^{D-2} \int\sqrt{^{D}g}\,d^{D}x\,(L_g + L_m),
\nn
     L_g \eql F(R) + c_1 R^{AB}R_{AB} + c_2 \cK,
\ear
  where $F(R)$ is an
  arbitrary smooth function of the $D$-dimensional scalar curvature $R$,
  $c_1$ and $c_2$ are
  constants, $L_m$ is a matter Lagrangian and ${^D}g = |\det(g_{MN})|$.

  We suppose that $b_{ab}$ describes a compact $d_1$-dimensional space of
  nonzero constant curvature, i.e., a sphere ($K=1$) or a compact
  $d_1$-dimensional hyperbolic space ($K = -1$) with a unit curvature
  radius; we use the system of units in which the speed of light $c$, the
  Planck constant $\hbar$ and the $D$-dimensional Planck length are equal to
  unity. Thus all quantities are now expressed in ($D$-dimensional)
  Planck units.

  The field equations of the full theory (\ref{act1}) are very complicated.
  Let us simplify the theory in the following way:

\medskip\noi
{\bf (a)}
    Integrate out the extra dimensions and express everything in terms of 4D
    variables and $\beta(x)$; we have, in particular,
\bearr                                                      \label{R4}
        R = R_4 + \phi + f_1,
\nnn
        f_1 = 2d_1 \Box \beta + d_1(d_1+1)(\d{\beta})^2,
\ear
    where $R_4$ is the 4D scalar curvature, $(\d{\beta})^2 = g\MN \d_\mu\beta
    \d_\nu\beta$, and we have introduced the effective scalar field
\beq                                                         \label{phi}
        \phi (x) = R_b \e^{-2\beta (x)}
              = K d_1(d_1-1)\, \e^{-2\beta (x)}   
\eeq
    The sign of $\phi$ coincides with $K = \pm 1$, the sign of curvature in
    the $d_1$ extra dimensions.

    The action in four dimensions has the form
\beq
     S = \Half \cV [d_1]\,\int\sqrt{^4g}\,d^{4}x\,
            \e^{d_1\beta}\,[L_g + L_m],                      \label{act2}
\eeq
    where $^{4}g = |\det(g\mn)|$ and $\cV[d_1 ]$ is the volume of a
    compact $d_1$-dimensional space of unit curvature.

\medskip\noi
{\bf (b)} Suppose that all quantities are slowly varying, i.e., consider
    each derivative $\d_{\mu}$ as an expression containing a small parameter
    $\eps$; neglect all quantities of orders higher than $O(\eps^2)$ (see
    \cite{VfromD, BRook}).

\medskip\noi
{\bf (c)} Perform a conformal mapping leading to the Einstein conformal
    frame, where the 4-curvature appears to be minimally coupled to the
    scalar $\phi$.

\medskip
  In the decomposition (\ref{R4}), both terms $f_1$ and $R_4$ are regarded
  small in our approach, which actually means that all quantities,
  including the 4D curvature, are small as compared with the
  $D$-dimensional Planck scale. The only term which is not small is
  $\phi$, and we can use a Taylor decomposition of the function $F(R) =
  F(\phi + R_4 + f_1)$:
\bearr                                                       \label{Fapprox}
    F(R) = F(\phi + R_4 + f_1 )
\nnn \cm
    \simeq F(\phi) + F'(\phi)\cdot(R_4 +f_1 )+...,
\ear
  with $F'(\phi)\equiv dF/d\phi$. In (\ref{act2}), we obtain, up to
  $O(\eps^2)$,
\bearr
      L_g = F'(\phi) R_4  + F(\phi) + F'(\phi) f_1 + c_*\phi^2
\nnn \cm
      + 2 c_1\phi \Box\beta
            + 2(c_1 d_1 + 2c_2) (\d\beta)^2                \label{Lg_4}
\ear
  with  $c_* = c_1/d_1 + 2c_2/[d_1(d_1-1)]$.

  The action (\ref{act2}) with (\ref{Lg_4}) is typical of a scalar-tensor
  theory (STT) of gravity in a Jordan frame. To study the dynamics of the
  system, it is helpful to pass on to the Einstein frame. Applying the
  conformal mapping
\bearr  \nhq                                                \label{trans-g}
    g\mn \mapsto \tg\mn = |f(\phi)| g\mn,
\ \ \
            f(\phi) =  \e^{d_1\beta}F'(\phi),
\ear
  after a lengthy calculation, we obtain the action in the Einstein frame as
  \cite{VfromD, we-13}
\bear
     S \eql \Half \cV[d_1] \int \sqrt{\tg}\, (\sign F') L,
\nn
     L \eql \tR_4 + K_{\rm E}(\phi) (\d\phi)^2
                        - 2V_{\rm E}(\phi) + {\wt L}_m,      \label{Lgen}
\\
     {\wt L}_m \eql (\sign F')\frac{\e^{-d_1\beta}}{F'(\phi)^2} L_m;
                                 \label{Lm}
\\ \nq
      K_{\rm E}(\phi) \eql                                   \label{KE}
        \frac{1}{4\phi^2} \biggl[
            6\phi^2 \biggl(\frac{F''}{F'}\biggr)^2\!
            -2 d_1 \phi \frac{F''}{F'}
\nnn \cm
        + \Half d_1 (d_1{+}2) + \frac{4(c_1 + c_2)\phi}{F'}\biggr],
\\ \nq
       -2V_{\rm E}(\phi) \eql (\sign F') \frac{\e^{-d_1\beta}}{F'(\phi)^2}
                [F(\phi) + c_* \phi^2],                      \label{VE}
\ear
  where the tilde marks quantities obtained from or with $\tg\mn$; the
  indices are raised and lowered with $\tg\mn$; everywhere $F = F(\phi)$ and
  $F' = dF/d\phi$; $\e^{\beta}$ is expressed in terms of $\phi$ using
  (\ref{phi}).

  Let us consider the electromagnetic field $F\mn$ as matter in the initial
  Lagrangian, putting
\beq                                                        \label{Lem-1}
               L_m = \alpha_1^{-1} F\mn F\MN,
\eeq
  where $\alpha_1$ is a constant. After reduction to four dimensions this
  expression acquires the factor $\e^{d_1\beta}$ arising from the metric
  determinant: $\sqrt{^D g} = \sqrt{^4 g} \e^{d_1\beta}$. In the subsequent
  transition to the Einstein picture the expression $\sqrt{^4 g}F\mn F\MN $
  remains the same (due to conformal invariance of the
  electromagnetic field), hence the Lagrangian (\ref{Lm}) takes the form
\beq
             {\wt L}_m = \alpha_1^{-1} \e^{d_1\beta} F\mn F\MN,
\eeq
  and for the effective fine structure constant $\alpha$ we obtain
\beq                                                         \label{var-A}
             \frac{\alpha}{\alpha_0} = \e^{d_1(\beta_0 - \beta)},
\eeq
  where $\alpha_0$ and $\beta_0$ are values of the respective quantities
  at a fixed space-time point, for instance, where and when the observation
  is taking place.

\section {Isotropic cosmologies}
\def\od{{\overline d}}

\subsection {Equations for small $\phi$}

  Depending on the choice of $F(R)$, the constants $c_1$ and $c_2$ and the
  matter Lagrangian, the theory (\ref{act1}) can lead to a great variety of
  cosmological models. Some of them were discussed in \cite{VfromD}, mostly
  those related to minima $V = V_{\min}$ of the effective potential
  (\ref{VE}) at nonzero values of $\phi$. Such minima correspond to
  stationary states of the scalar $\phi$, and consequently of the volume
  factor of extra dimensions that determines the effective FPC values. If
  $V_{\min} >0$, it can play the role of a cosmological constant that
  launches an accelerated expansion of the Universe.

  Here, as in \cite{we-13}, we focus on another minimum of the potential
  $V_{\rm E}$, existing for generic choices of the function $F(R)$ with
  $F'> 0$ and located at $\phi =0$. If $\phi \to 0$ at late times, this
  corresponds to growing rather than stabilized extra dimensions: $b =
  \e^{\beta}\sim 1/\sqrt{|\phi|} \to \infty$. Such a model can still be of
  interest if the growth is sufficiently slow and the size $b$ does not
  reach detectable values by now. We can recall that the admissible range of
  such growth comprises as many as 16 orders of magnitudes if the
  $D$-dimensional Planck length $l_D = 1/\mD$ coincides with the 4D one,
  i.e., about $10^{-33}$ cm: the upper bound corresponds to lengths about
  $10^{-17}$ cm or energies of the order of a few TeV. This estimate
  certainly changes if there is no such coincidence.

  Assuming small values of $\phi$, we should still take care of not
  violating our general assumptions, namely, the requirement that $|\phi|$
  is still very large as compared to 4D quantities. It is really so since
\[
       |\phi| = \frac{d_1(d_1 - 1)}{b^2},
\]
  where $b \lesssim 10^{16}$, hence $|\phi| \gtrsim d_1^2\cdot 10^{-32}$,
  whereas the quantity $\tR_4$, if identified with the curvature of the
  modern Universe, is $\sim 10^{-122}$ in Planck units (that is,
  close to (the Hubble time)${}^{-2}$, see also \eq (\ref{Hubble}) below.

  Let us check whether it is possible to describe the modern stage of
  the Universe evolution by an asymptotic form of the solution for small
  $\phi$ as a spatially flat cosmology with the 4D Einstein-frame metric
\beq
        d{\wt s}{}^2_4 = dt^2 - a^2 (t) d\vec x{}^2,         \label{dsE}
\eeq
  where $a(t)$ is the Einstein-frame scale factor.
  At small $\phi$, assuming a smooth function $F(\phi)$, we can restrict
  ourselves to the first three terms of its Taylor decomposition:\footnote
    {We assume for certainty $\phi > 0$, hence by (\ref{phi}), $K = +1$,
    but everything can be easily reformulated for $\phi < 0$.
     \label{foot-phi+}}
\beq                                                         \label{F_0}
      F(\phi) = -2\Lambda_D + F_1\phi + F_2\phi^2,
\eeq
  where $\Lambda_D$ is the initial cosmological constant. For simplicity, we
  suppose\footnote
    {The theory is insensitive to multiplying the action by a
     constant, and we use this freedom to fix $F_2=1$.
     \label{foot-F2}}
  $F_1 = 0$, $F_2 =1$. It is then convenient to rewrite the Lagrangian
  (\ref{Lgen}) at small $\phi$ in terms of $\beta$ instead of $\phi$:
\beq
    L = \tR_4 + 2K_0 (\d\beta)^2 - 2V(\beta) + {\wt L}_m,    \label{L-bet}
\eeq
  with
\beq                                                         \label{K0}
        K_0 = \frac{1}4{}\bigl[d_1^2 - 2d_1 + 12 + 4(c_1+c_2)\bigr].
\eeq
  Neglecting the gravitational influence of the electromagnetic field
  (that is, considering only vacuum models), one can write down the
  independent components of the Einstein and scalar field equations
  with the unknowns $\beta(t)$ and $a(t)$ as follows:
\bearr
       3\frac{\dot a{}^2}{a^2}                                \label{eq-a4}
            = K_0 \dot\beta{}^2 + V(\beta),
\yyy                                                            \label{eq-b4}
      2K_0\Bigl(\ddot \beta + 3\frac{\dot a}{a} \dot{\beta}\Bigr)
                = - \frac{dV}{d\beta}.
\ear

  The form of the potential $V(\beta)$ depends on further assumptions.
  If $\Lambda_D \ne 0$, we have
\bearr                                                       \label{V1}
        V = V_1(\beta) = V_{10} \e^{-2\od \beta},
\nnn
     V_{10} = \frac{\Lambda_D}{4 d_1^2(d_1-1)^2}, \cm \od = \frac{d_1-4}{2}.
\ear
  and $d_1 > 4$ is required here. This model was considered in \cite{we-13},
  and we here discuss it for comparison with another model, in which
  $\Lambda_D = 0$ (or, more generally, $\Lambda_D \ll \phi^2$).

  Under the assumption $\Lambda_D = 0$ we have
\bearr                                                        \label{V2}
        V = V_2(\beta) = V_{20} \e^{-d_1\beta},
\nnn
    V_{20} = -\frac{1 + c_*}{8}
           = -\frac 18 \biggl[ 1 + \frac{c_1}{d_1}
                + \frac{c_2}{d_1(d_1-1)}\biggr].
\ear

\subsection{Model 1: $F(R) = -2\Lambda + R^2$}

  \eqs (\ref{eq-a4}) and (\ref{eq-b4}) correspond to a scalar field with
  an exponential potential and can be solved exactly, but the solution looks
  rather involved, and for our purpose more preferable is the comparatively
  simple approximate solution obtainable in the slow-rolling approximation
  that should be acceptable at late times. Let us suppose that
\beq                                                       \label{rolling}
       |\ddot \beta| \ll 3\frac{\dot a}{a} \dot {\beta},
   \qquad
        K_0 \dot\beta{}^2 \ll V(\beta),
\eeq
  and neglect the corresponding terms in \eqs (\ref{eq-a4}) and
  (\ref{eq-b4}). Then, expressing the quantity $\dot a/a$ from (\ref{eq-a4})
  and substituting it into (\ref{eq-b4}) with $V = V_1$, we obtain
  an expression for $\dot\beta$ whose integration gives
\beq                                                        \label{sol-b}
    \e^{\od\beta} = \frac{\od^2}{K_0}\sqrt{\frac{V_{10}}3}(t+t_1),
\eeq
  where $t_1$ is an integration constant. For the scale factor
  $a(t)$ we have ${\dot a}/a = p/(t + t_1)$ whence
\beq                                                        \label{sol-a}
        a =  a_1 (t+t_1)^p,  \quad\
        a_1 = \const,   \quad\  p = \frac {K_0}{\od^2}.
\eeq

  One can verify that the slow-rolling conditions (\ref{rolling}) hold as
  long as $3p \gg 1$, or in terms of the input parameters of the theory,
\beq                                                         \label{roll}
       3p = 3\frac{d_1^2 -2 d_1 +12 + 4(c_1+c_2)}{(d_1-4)^2} \gg 1.
\eeq
  Let us assume that this condition holds.

  A further interpretation of the results depends on which conformal
  frame is regarded physical (observational) \cite{bm-predict,erice},
  and this in turn depends on the manner in which fermions appear in the
  (so far unknown) underlying unification theory involving all interactions.

  Let us make some estimates assuming that the observational picture is
  Einstein's. The inverse of the modern value of the Hubble parameter (the
  Hubble time) is estimated as\footnote
    {As usual in cosmology, here and henceforth the subscript ``0''
    refers to the present epoch.}
\beq                                                        \label{Hubble}
    t_H = \frac{1}{H_0} = \frac{a_0}{{\dot a}_0} \approx
    4,4 \times 10^{17}\, {\rm s}
                \approx 8 \times 10^{60}\, t_{\rm pl},
\eeq
  where $t_{\rm pl}$ is the Planck time.
  From (\ref{sol-a}) it follows that $H_0 = p/(t_0 + t_1)$, whence
\beq
       t_* := t_0 + t_1 = p t_H \gg t_H.
\eeq
  With $p \gg 1$, the power-law expansion is close to exponential, and
  the model satisfies the observational constraints on the factor $w$ in the
  effective equation of state $p = w\rho$ of dark energy: at $w = \const$ we
  have $a \sim t^{2/(3+3w)}$, consequently, $w = -1 + 2/(3p)$ is a number
  close to $-1$.

  The ``internal'' scale factor $b(t)= \e^\beta$ grows much
  slower than $a(t)$:
\beq                                    \label{b_t}
        b(t) = b_0 \Bigl(\frac{t+t_1}{t_*}\Bigr)^{1/\od}\!, \ \
    b_0 = \biggl(\!\frac 1{H_0} \sqrt{\frac {V_1}{3}}\biggr)^{1/\od}.
\eeq

  Using the expression for $V_{10}$ in (\ref{V1}), one can estimate the
  initial parameter $\Lambda_D$ in terms of the present size $b_0$ of the
  extra factor space: in Planck units,
\bearr                                                  \label{Lam-est}
       \Lambda_D = 12 H_0^2 d_1^2 (d_1-1)^2 b_0^{d_1-4}
\nnn \cm
        \approx \frac {3}{16} d_1^2 (d_1-1)^2 b_0^{d_1-4} \times 10^{-120}.
\ear
  As already mentioned, $b_0$ should be in the range $1 \ll b_0 \lesssim
  10^{16}$ in Planck units. The estimate (\ref{Lam-est}) shows that the
  present model makes much easier the well-known ``cosmological constant
  problem'' (the difficulty of explaining why in standard cosmology
  $\Lambda_{\rm standard} \sim 10^{-122}$ in Planck units). For example, if
  (in the admissible range) $b_0 = 5 \ten{14}$ and $d_1 = 12$, it
  follows $\Lambda_D \approx 12.76$, without any indication of fine tuning.

  Other initial parameters, $c_1$ and $c_2$, should not be too large: as
  estimated in \cite{we-13}, they should not exceed $10^{11}$, otherwise
  our basic assumptions can be violated. But the smallness of the observed
  variations of $\alpha$ indicates that they should not be too small.
  Indeed, according to (\ref{var-A}),
\beq       \nq
    \frac{\alpha}{\alpha_0}
        = \biggl(\frac{t+t_1}{t_0+t_1}\biggr)^{\!-2d_1/(d_1-4)} \!
                \approx 1 - \frac{2d_1}{d_1-4}\frac {t-t_0}{t_*},
\eeq
  so that $\dot \alpha/\alpha \sim 10^{-10}/p$ per year. By the empirical
  data, this quantity cannot be larger than about $10^{-17}$ per year.
  This leads to the constraint $p \gtrsim 10^7$. The allowed range of $c_1$
  and $c_2$ (assuming that they are both positive and
  have the same order of magnitude)
\beq                            \label{ineq1}
            10^6 <  c_{1,2} \ll 10^{11}
\eeq
  is wide enough, and there is no fine tuning. Moreover, we shall see that
  the inequality $c_{1,2} > 10^6$ is substantially relaxed in the perturbed
  model.

\subsection{Model 1 in Jordan's frame}

  It is of interest how the same model looks in the Jordan frame
  corresponding to the initial D-dimensional action. To obtain it, we return
  to the transformation (\ref{trans-g}) and find
\beq                                                           \label{jo1}
    ds^2_{\rm J} = ds^2_{\rm E}/f, \cm f = \e^{d_1\beta}F'(\phi).
\eeq
  Since $F' = 2\phi \sim \e^{-2\beta}$, with the solution (\ref{b_t}) we have
\beq
    f = \const \cdot (t + t_1)^{2(d_1-2)/(d_1-4)}.            \label{f-jo}
\eeq
  We apply this transformation to the cosmological metric (\ref{dsE}),
  putting
\beq                                                          \label{ds-jo}
    ds_{\rm J} = d\tau^2 - a^2_{\rm J}(\tau) d\vec x{}^2
                = \frac{1}{f} [dt^2 - a^2(t) d\vec x{}^2],
\eeq
  where $\tau$ and $a_{\rm J}(\tau)$ are the cosmological time and the scale
  factor in the Jordan frame, respectively. Then $d\tau= dt/\sqrt{f}$ and
  $a_{\rm J}(\tau) = a(t)/\sqrt{f}$. Integrating, we find
\beq                                                          \label{tau1}
    \tau - \tau_1 = -\od\, t_* ^{1 + 1/\od}(t+t_1)^{-1/\od},
            \ \ \ \tau_1 = \const.
\eeq
  Substituting (\ref{tau1}) into the solution (\ref{sol-a}), (\ref{b_t}), we
  obtain
\bear                                                         \label{a_tau}
    a_{\rm J}(\tau) \eql a_{1*} (\tau_1 - \tau)^{-\od p + \od +1},
        \ \  a_{1*} = \const,
\yy                                                           \label{b_tau}
    b(\tau) \eql b_0 \frac{\tau_1 - \tau_0}{\tau_1 - \tau},
\ear
  where $b_0$ and $\tau_0$ are the present-time values of $b$ and $\tau$.
  It is a big-rip cosmology: both scale factors $a$ and $b$ blow up at a
  certain time $\tau_1$ in the future. Moreover, comparing the behavior of
  $a_{\rm J}(\tau)$ with the dependence $a \sim \tau^{2/(3+3w)}$,
  corresponding to a model with a constant-$w$ perfect fluid ($p = w\rho$),
  we see that now the effective equation-of-state parameter is
\beq
    w = -1 - \frac{2}{3}(\od p -\od - 1).           \label{w_J}
\eeq
  The solution (\ref{sol-a}), (\ref{b_t}) is valid under the assumption $p
  \gg 1$, therefore the parameter (\ref{w_J}) is much smaller than $-1$, and
  we have to conclude that the Jordan-frame version of Model 1
  contradicts the observations.

  We conclude that this model can be viable only if the Einstein frame is
  interpreted as the observational one.

\subsection{Model 2: $F(R) = R^2$}

  Let us now try to build a cosmological model with a purely quadratic
  function $F(R)$, that is, $F(R) = R^2$ (see footnote \ref{foot-F2}).
  As before, the Einstein-frame 4D scale factor $a(t)$ and the effective
  scalar field $\beta(t)$ obey the same equations (\ref{eq-a4}) and
  (\ref{eq-b4}), but with $V_1$ replaced by $V_2$. For the model to be
  viable, we need a positive potential $V$ and therefore assume
  $V_{20} >0$ which is possible under a proper choice of $c_1$ and $c_2$.

  The solution in the slow-rolling approximation (\ref{rolling}), obtained
  in the same way as before, has the form
\bearr                                                     \label{sol-a2}
    a(t) = a_2(t +t_2)^q,
\nnn \cm
    q= \frac{1}{d_1^2}[d_1^2 -2d_1 +12 +4 (c_1+c_2)];
\yyy\nq\,                                                   \label{sol-b2}
    b(t) \equiv \e^\beta = b_0 \biggl(\frac{t+t_2}{t_{**}}\biggr)^{\! 2/d_1}
    \nq,    \ \ \
    b_0 = \biggl(\frac{t_{**}\sqrt{V_{20}}}{\sqrt{3}q}\biggr)^{\! 2/d_1}\nq,
\nnn
\ear
  where $a_2$ and $t_2$ are integration constants and $t_{**} = t_0 + t_2$,
  $t_0$ being the present time. Identifying the present Hubble parameter
  $H_0$ with the present value of $\dot a/a$ according to (\ref{sol-a2}), we
  obtain $t_* = 1/H_0 = q t_H$, where $t_H$ is the Hubble time. And, as
  before, we easily verify that the slow-rolling conditions (\ref{rolling})
  hold as long as $q \gg 1$. Under this condition the model adequately
  describes the present state of the Universe since the effective
  equation-of-state parameter is $w = -1 + 2/(3q)$, as in Sec.\,3.2.

  This model does not contain an initial cosmological constant like
  $\Lambda_D$, and instead of the estimate (\ref{Lam-est}) we have a
  constraint on $d_1$ that follows from the expression for $b_0$ in
  (\ref{sol-b2}). Indeed, since $t_{**} = q/H_0$, we have
\beq                                                            \label{b0-d}
    b_0 = \biggl(\frac{\sqrt{V_{20}}}{\sqrt{3}H_0}\biggr)^{2/d_1}
        \lesssim 10^{16},
\eeq
  while (see (\ref{Hubble}) $1/H_0 = t_H \sim 10^{60}$ in Planck units,
  hence, assuming that $V_{20} \sim 1$, we must have $d_1 \geq 8$.

  The constraints on the input parameters $c_1$ and $c_2$ are now different
  from (\ref{ineq1}) because, while the condition $q \gg 1$ requires
  $c_1 + c_2 \gg 1$, to have $V_{20} \sim 1$, we must require $c_* < -1$.

  For time variations of $\alpha$ we now have
\bearr   \nq
    \frac{\alpha}{\alpha_0} = \biggl(\frac{b}{b_0}\biggr)^{-d_1}\!\!
        = \biggl(\frac{t+t_2}{t_*}\biggr)^{-2}\!\!
        \approx 1 - 2\frac{t-t_0}{t_*},                \label{A2}
\nnnv
    \dot\alpha/\alpha_0 \approx -2H_0/q = -2/(qt_H).
\ear
  The empirical bounds on $\dot\alpha$ require $q \gtrsim 10^7$, which leads
  to the constraint $c_1 + c_2 \gtrsim 10^6$. This can be combined with
  $c_* < -1$ if $c_2 \gtrsim 2\ten{6}$ and $c_1 \lesssim -10^6$. And, as
  in the first model, these conditions are substantially relaxed for the
  perturbed configuration, see Section 4.

\subsection {Model 2 in Jordan's frame}

  In the transformation (\ref{trans-g}) we now have
\beq
    ds^2_{\rm J} = \frac{ds^2_{\rm E}}{f(\phi)},\ \ \
            f(\phi) = \const\cdot (t + t_2)^{2-4/d_1}.
\eeq
  Applying this transformation to the metric (\ref{dsE}), we once again
  obtain \eq (\ref{ds-jo}) where $\tau$ and $a_{\rm J}(\tau)$ are the
  cosmological time and the scale factor in Jordan's frame, respectively.
  Integrating $d\tau= dt/\sqrt{f}$ and substituting it to $a_{\rm J}(\tau)=
  a(t)/\sqrt{f}$, we find
\bear \nq                               \label{tau2}
    \tau + \tau_2 \eql \!\int\! \frac{dt}{(t+t_2)^{1-2/d_1}}
            \sim (t + t_2)^{2/d_1},
\yy                                                          \label{a_tau2}
    a_{\rm J}(\tau) \eql a_{2*} (\tau + \tau_2)^s,\ \
            s = \half d_1(q-1) + 1;
\yy                                                           \label{b_tau2}
    b(\tau) \eql b_0 \frac{\tau + \tau_2}{\tau_*}, \ \ \
            \tau_* = \tau_0 + \tau_2,
\ear
  where $\tau_2$ and $a_{2*}$ are integration constants, $b_0$ and $\tau_0$
  are the present-day values of $b$ and $\tau$, and $b_0$ is still given by
  (\ref{b0-d}) leading to the same constraint $d_1\geq 8$.

  The viability of this Jordan picture as an accelerated cosmology is
  provided by the condition $s > q \gg 1$ in (\ref{a_tau}). We should also
  consider time variations of $\alpha$ and the gravitational constant
  $G$: the latter is constant by definition in the Einstein frame but
  changes in Jordan's by the same law as $\alpha$,
\beq                                                          \label{G-dot2}
    G/G_0 = \alpha/\alpha_0 = \e^{d_1(\beta_0-\beta)}.
\eeq
  Variations of $\alpha$ on Earth are restricted to the 17th significant
  digit, see (\ref{a-clock}), (\ref{a-Oklo}). The ``Australian dipole''
  (\ref{dipole}) testifies to variations on the level of $10^{-15}$
  per year. The observational constraints on $G$ are much weaker, see
  (\ref{G-dot}). Consequently, if we constrain the parameters of our model
  by variations of $\alpha$, the observational bounds on $G$ variations hold
  automatically and need not be considered separately.

  The Hubble parameter is here $H_0 = s/\tau_*$, for $d\alpha/d\tau$
  we have the expression $\alpha_0 (d_1/s)H_0$, and since $d_1/s \approx 2/q$
  for $d_1\geq 8$, the estimate (\ref{A2}) still remains valid and leads to
  the same constraints on $c_1$ and $c_2$ as in the Einstein frame.

\section {$x$-dependent perturbations and varying $\alpha$}

\subsection {General relations}

  We have discussed the properties of homogeneous models which cannot
  account for spatial variation of any physical quantity.

  Let us now try to describe variations of $\alpha$ by taking into account
  spatial perturbations of the effective scalar field and the metric. We
  take the Einstein-frame metric more general than (\ref{dsE}),
\beq                                                         \label{ds1}
     ds_{\rm E}^2 =
    \e^{2\gamma}dt^2 - \e^{2\lambda}dx^2 -\e^{2\eta}(dy^2 + dz^2),
\eeq
  where $\gamma, \lambda, \eta$ are functions of $x$ and $t$. We will not
  discuss the reasons why the metric perturbation has a distinguished
  direction, only mentioning a possible weak inhomogeneity at the beginning
  of the inflationary period and the opportunity of domain walls
  that can be thick on the cosmological scale.

  The conditions that the metric (\ref{ds1}) only slightly differs from
  (\ref{dsE}) are
\bearr
    \gamma = \delta\gamma(x,t),  \ \
    \lambda = \ln a(t) + \delta\lambda (x,t),
\nnn \cm
    \eta = \ln a(t) + \delta\eta (x,t),
\earn
  where all ``deltas'' are small. We also replace the effective scalar field
  $\beta (t)$ with $\beta(t) + \delta\beta(x,t)$.  Then the relevant
  Einstein-scalar equations corresponding to the Lagrangian (\ref{L-bet})
  can be written as follows (preserving only terms linear in the ``deltas''):
\bearr
    \delta\ddot{\beta} + \frac{3\dot a}{a} \delta\dot{\beta}
    + \dot\beta (\delta\dot\lambda - \delta\dot \gamma)
    - \frac{\delta \beta''}{a^2}                                \label{sc-d}
    + \frac{\delta(V_\beta \e^{2\gamma})}{2K_0} =0,
\nnn
\\ \lal                                                         \label{22-d}
     \frac{\dot a}{a}(\delta\dot\lambda - \delta\dot \gamma)
            = \delta(V \e^{2\gamma}),
\\ \lal                                                         \label{01-d}
     \frac{\dot a}{a} \delta\gamma' = K_0 \dot\beta \, \delta\beta',
\ear
  where we choose the gauge (i.e,, the reference frame in perturbed
  space-time) $\delta \eta \equiv 0$, dots and primes stand for $\d/\d t$
  and $\d/\d x$, respectively. We also denote $V_\beta = dV/ d\beta$.

  Without loss of generality, (\ref{01-d}) leads to
\beq                                                          \label{d_gam}
        \delta\gamma = \frac{K_0}{H} \dot{\beta} \delta\beta,
\eeq
  where, as before, $H = \dot a/a$. Substituting this $\delta\gamma$ to
  (\ref{sc-d}) and taking the difference $\delta\dot\lambda - \delta\dot
  \gamma$ from (\ref{22-d}), we arrive at the following single wave equation
  for $\delta\beta$:
\bearr                                                         \label{eq-db}
    \delta \ddot \beta + \frac{3\dot a}{a} \delta\dot\beta
        -\frac{1}{a^2} \delta\beta''
\nnn \ \ \
     + \delta\beta \biggl[ \frac{2 \dot\beta{}^2}{H^2} VK_0
    + \frac{2 \dot\beta}{H} V_\beta + \frac 1{2K_0}V_{\beta\beta}\bigg]=0
\ear
  with an arbitrary constant $K_0$ and an arbitrary potential $V (\beta)$.
  Assuming that the background quantities $a(t)$ and $\beta(t)$ are known,
  it remains to find a solution for $\delta\beta$ which, being added to the
  background $\beta(t)$, would be able to account for the observed picture
  of variations of $\alpha$.

  Since the background is $x$-independent, we can separate the
  variables and assume
\[
    \delta\beta = y(t) \sin k(x+x_0)
\]
  where $k$ has the meaning of a wave number. Then $y(t)$ obeys the equation
\bearr                                                       \label{eq-y}
    \ddot y + \frac{3\dot a}{a} y + \frac{k^2 y}{a^2}
\nnn \ \ \
     + \biggl[ \frac{2 \dot\beta{}^2}{H^2} VK_0
    + \frac{2 \dot\beta}{H} V_\beta + \frac 1{2K_0}V_{\beta\beta}\biggr]y=0
\ear

  Since \eq (\ref{eq-y}) is itself approximate and describes only a
  restricted period of time close to the present epoch. it is reasonable to
  seek its solution in the form of a Taylor series near $t = t_0$,
\beq
      y(t) = y_0 + y_1 (t-t_0) + \Half y_2 (t-t_0)^2 + \ldots,
\eeq
  with $y_i = \const$. Then $y_0$ and $y_1$ can be fixed at will as initial
  conditions. Even more than that, for a certain neighborhood of $t=t_0$ we
  can simply suppose $y = y_0 + y_1 (t-t_0)$. Actually, in our models this
  approximation is good enough for $t-t_0 \ll t_*$.

  In this approximation we obtain the following expression for
  variations of $\alpha$:
\bearr                                                      \label{dA1}
      \frac{\alpha}{\alpha_0} \approx 1 + d_1 [\beta_0 - \beta(t)
\nnn \ \
    - (y_0 + y_1(t-t_0)) \sin k(x+x_0)] + O(\ep^2),
\ear
  where $O(\ep^2)$ means $O((t-t_0)^2/t_*^2)$ and $\beta(t)$ is the
  background solution in a particular model.

\subsection {The two models in Einstein's frame}

{\bf Model 1.} The function $\beta(t)$ is given by (\ref{b_t}).
  Assuming that the observer is located at $x=0$ and requiring
  $\alpha/\alpha_0 = 1 + O(\ep^2)$ at $x=0$, we obtain the condition
\beq
     y_1 \sin (kx_0) = - 1/(\od \,t_*).                    \label{y_1}
\eeq
  This explains very small, if any, variations of $\alpha$ on Earth at
  present and since the Oklo times. To account for the ``Australian
  dipole'', we need an approximately linear dependence $\alpha(x)$
  on the past light cone of the point $t=t_0$, $x=0$.

  At small enough $k$ (that is, assuming a very long wave of perturbations),
  so that $kx \ll 1$ and $kx \sim (t-t_0)/t_*$, a substitution of (\ref{y_1})
  into (\ref{dA1}) leads to \cite{we-13}
\bearr
       \alpha/\alpha_0 \approx 1 - d_1 y_0 \sin (kx_0)  \label{dA2}
                   + d_1 y_0 \,kx \cos (kx_0)
\nnn \inch
           + O(\ep^2),
\ear
  so that a time dependence is eliminated from the expression for $\alpha$
  up to $O(\ep^2)$ while the $x$ dependence is linear, as required.
  (Though, since the measurement errors are rather large,
  the $x$ dependence should not necessarily be strictly linear.)

\begin{figure}[ht]
\centering
\vspace{-7cm}
\includegraphics[scale=0.45]{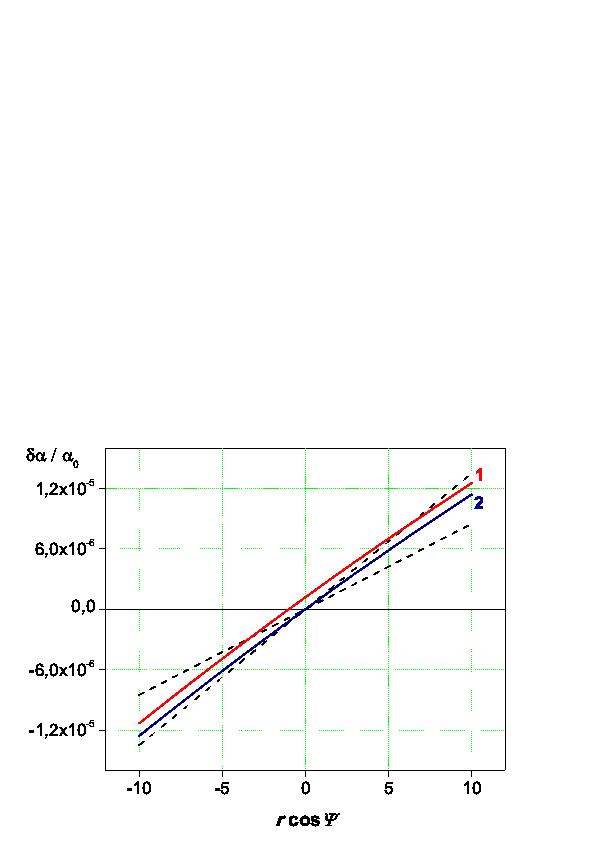}
\caption{\small The $r$ dependence of $\delta\alpha/\alpha_0$; the distance
    $r$ is measured in billions of light years (BLY) (from \cite{we-13}).
    The dashed lines correspond to \eq (\ref{dipole}), the solid lines to
    \eq (\ref{dA1}) at the parameter values $d_1 = 12$, $p = 10^{7}$, $y_0 =
    -2\ten{-5}$, $y_1 = -10^{-7}$ (bill. years)$^{-1}$, $k = 0.005$
    (BLY)$^{-1}$. Line 1 corresponds to $x_0 = 1$ BLY, line 2 to $x_0 = 0.01$
    BLY.}
    \label{picture}
\end{figure}
  This model contains the input theoretical parameters $d_1,\, c_1,\ c_2$,
  and the constants $k$, $x_0$, $y_0$, $y_1$, which can be ascribed to
  slightly inhomogeneous initial conditions at primordial inflation.
  Their choice enables us to explain the spatial variations of
  $\alpha$ in agreement with the observations \cite{webb2}. Actually, there
  are only two conditions imposed on them: (\ref{y_1}) and the relationship
  identifying (\ref{dA2}) with the expression (\ref{dipole}) at $r=x$ and
  $\cos\psi =1$, i.e., on the dipole axis. We obtain (in Planck units)
\beq
    d_1 y_0 k \cos (kx_0) \approx -2 \times 10^{-66}.
\eeq
  The small constant shift of the $\alpha$ value at $x=0$ against the
  background does not change the interpretation of these results.

  The constraint on the input parameters $c_1$ and $c_2$ due to slow
  variations of $\alpha$ on Earth is now cancelled since this condition is
  provided by the equality (\ref{y_1}). We should only provide the
  validity of the approximation in which we work, that is, $p \gg 1$; it
  holds fairly well if $c_1 + c_2 \gtrsim 1000$. Hence the inequality
  (\ref{ineq1}) is replaced by a much weaker one:
\beq                                    \label{ineq2}
        1000 \lesssim c_1 \sim c_2 \ll 10^{11}.
\eeq

  Fig.\,2 presents the distance dependence of $\delta\alpha/\alpha$
  (see (\ref{dA1})) for some values of the parameters.

\medskip\noi
{\bf Model 2.} Now the background function $\beta(t)$ is given by
  (\ref{sol-b2}), and as a result,
\bearr                                                      \label{dA1'}
      \frac{\alpha}{\alpha_0} \approx 1
            - d_1 \sin[k(x+x_0)]\,[y_0 + y_1(t-t_0)]
\nnn \inch
        - \frac{2(t-t_0)}{t_*} + O(\ep^2),
\ear
  To eliminate variations of $\alpha$ at $x=0$, we require
\beq
     y_1 \sin (kx_0) = - 2/(d_1\,t_*).                    \label{y_1'}
\eeq
  Then, for $x \ll t_*$, and $kx \ll 1$, we obtain on the past light cone
  an expression coinciding with (\ref{dA2}):
\bearr
       \frac{\alpha}{\alpha_0} \approx 1
       - d_1 y_0 \sin (kx_0)                \label{dA2'}
                   + d_1 y_0 \,kx \cos (kx_0).
\ear
  Fig.\,2, showing a comparison of our models with the Australian
  dipole (\ref{dipole}), thus applies to both kinds of models, with
  and without the term $\Lambda_D$ in $F(R)$ in the initial action.

\subsection{Model 2 in Jordan's frame}

  What changes in Jordan's frame if we use the same solution for
  $\beta(t,x)$? The conformal mapping does not affect the expression for
  $\alpha(x,t)$ as well as the light cone, only the scales along all
  coordinate axes change, causing slightly different relationships in
  terms of the Jordan-frame cosmic time $\tau$. We now assume
  $y = y_0 + y_1 (\tau-\tau_0)$ and obtain
\bearr \nq                                                 \label{dA1j}
      \frac{\alpha}{\alpha_0} = 1
        - d_1 [y_0 + y_1(\tau{-}\tau_0)\sin k(x{+}x_0)]
\nnn \inch
       -\frac{d_1(\tau{-}\tau_0)}{\tau_*} + O(\ep^2),
\ear
  Zero variations at $x=0$ are now provided by
\beq
    y_1 \sin (kx_0) = -1/\tau_*.
\eeq
  Under this condition, for $\tau-\tau_0 \ll \tau_*$, $kx \ll 1$, we have
  again the same relation (\ref{dA2}) or (\ref{dA2'}) for purely spatial
  variations of $\alpha$. Since it is the Jordan picture, the same
  variations are predicted for the gravitational constant $G$.

\section{Conclusion}

  Continuing the study begun in \cite{we-13}, we have considered some
  cosmological models that follow from curvature-nonlinear multidimensional
  gravity in the slow-change (on the Planck scale) approximation, which after
  reduction to four dimensions turn into multi-scalar-tensor gravity.
  We have shown that the recently reported variations of the fine structure
  constant $\alpha$ in space and time \cite{webb1, webb2} can be explained
  in the framework of such models chosen as slightly perturbed isotropic
  models: both an accelerated expansion and variations of $\alpha$ follow
  from the behavior of the scalar field of multidimensional origin.
  The agreement with observations is provided by the choice of initial data,
  which can be interpreted as random values of the extra-dimensional metric
  at the inflationary stage of the Universe. Thus spatial and temporal
  variations of $\alpha$ can be manifestations of the multidimensional
  space-time geometry.

  The models described here are very simple, approximate and tentative but
  still work fairly well at times not too far from the present epoch. They
  do not take into account other kinds of matter than dark energy; their
  inclusion must lead to a better description on a wider time interval.
  A possible existence of more than one extra factor space, which will lead
  to more than one effective scalar field in four dimensions, can hopefully
  lead to a viable model covering the whole classical evolution of the
  Universe beginning with primordial inflation.

  We showed that the model considered in \cite{we-13} is viable only in
  Einstein's conformal frame and suggested another model, viable in both
  Einstein's and Jordan's pictures. In the latter, along with $\alpha$, the
  gravitational constant $G$ varies at the same rate. Such variations are in
  agreement with the observational constraints, see (\ref{G-dot}).
  Let us note that the prediction of simultaneous variations of different
  physical constants due to a common cause, namely, variation of the size of
  extra dimensions, is a common feature of multidimensional theories. Thus,
  for example, a possible future discovery of $G$ variations qualitatively
  different from those of $\alpha$, can put to doubt not only models of the
  kind considered here but the whole paradigm of multidimensional gravity.
  On the contrary, a discovery of parallel evolution of different constants
  would be a strong argument in favor of extra dimensions.

\subsection*{Acknowledgments}

  We thank M.I. Kalinin for helpful discussions.


\small
\begin{thebibliography}{99}

\bibitem{webb1}
    J. K. Webb et al., Further evidence for cosmological evolution of the
    fine structure constant. Phys. Rev. Lett. {\bf 87}, 091301 (2001).

\bibitem{webb2}
    J. K. Webb et al., Evidence for spatial variation of the fine structure
    constant. Phys. Rev. Lett. {\bf 107}, 191101 (2011);  ArXiv: 1008.3907.

\bibitem{beren1}
    J. C. Berengut and V. V. Flambaum, Astronomical and laboratory searches
    for space-time variation of fundamental constants.
    J. Phys. Conf. Ser. {\bf 264}, 012010 (2011); Arxiv: 1009.3693.

\bibitem{rosbnd1}
    T. Rosenband et al., Observation of the 1S0>3P0 Clock Transition in
    $^{27}{\rm Al}^{+}$. Phys. Rev. Lett. {\bf 98}, 220801 (2007).

\bibitem{shlya}
    A. I. Shlyakhter,
    Direct test of the constancy of fundamental nuclear constants.
    Nature {\bf 260}, 340 (1976).

\bibitem{fujii}
    Y. Fujii et al. The nuclear interaction at Oklo 2 billion years ago.
    Nucl. Phys. {\bf B573}. 377 (2000).

\bibitem{chiba}
    T. Chiba. The constancy of the constants of Nature: Updates.
    Prog. Theor. Phys. {\bf 126}, 993--1019 (2011); ArXiv: 1111.0092.

\bibitem{Chiba-1}
    T. Chiba and M. Yamaguchi,
    Runaway domain wall and space-time varying $\alpha$.
    JCAP \textbf{1103}, 044 (2011); ArXiv: 1102.0105.

\bibitem{Olive-1}
    K. A. Olive, M. Peloso, and J.-P. Uzan, The wall of fundamental
    constants.  Phys. Rev. D \textbf{83}, 043509 (2011); ArXiv: 1011.1504.

\bibitem{Olive-2}
    K. A. Olive, M. Peloso, and A. J. Peterson, Where are the walls?
    ArXiv: 1204.4391.

\bibitem{Odintsov}
    K. Bamba, S. Nojiri, S.D. Odintsov,
    ArXiv: 1107.2538.

\bibitem{Barrow}
    J. D. Barrow and S. Z. W. Lip,
    A generalized theory of varying alpha.
    ArXiv: 1110.3120.

\bibitem{Mariano}
    A. Mariano and L. Perivolaropoulos,
    Is there correlation between fine structure and dark energy cosmic dipoles?
    ArXiv: 1206.4055.

\bibitem{Mari-2}
    A. Mariano and L. Perivolaropoulos,
    CMB maximum temperature asymmetry axis: alignment with other cosmic
    asymmetries.
    ArXiv: 1211.5915.

\bibitem{we-13}
    K. A. Bronnikov, V. N. Melnikov, S. G. Rubin, and I. V. Svadkovsky,
    Nonlinear multidimensional gravity and the Australian dipole,
    ArXiv: 1301.3098.

\bibitem{mel1}
    V. N. Melnikov, Multidimensional classical and quantum cosmology and
    gravitation. Exact solutions and variations of constants.
    In: Cosmology and Gravitation, ed. M. Novello, Editions Frontieres,
    Singapore, 1994, p. 147.

\bibitem{mel2}
    V. N. Melnikov, Gravity and cosmology as key problems of the millennium.
     In: {\it Albert Einstein Century Int. Conf.}, eds. J.-M. Alimi and A.
    Fuzfa, AIP Conf. Proc. 861, 2006, p. 109.

\bibitem{VfromD}
    K. A. Bronnikov and S. G. Rubin,
    Self-stabilization of extra dimensions.
    \PRD {73} 124019 (2006).

\bibitem{mueller}
    J. Mueller and L. Biskupek.
    Variations of the gravitational constant from lunar laser ranging data.
    Class. Quantum Grav. {\bf 24}, 4533--4538 (2007).

\bibitem{bm-predict}
    K. A. Bronnikov and V. N. Melnikov,
    On observational predictions from multidimensional gravity.
    \GRG {33} 1549 (2001).

\bibitem{erice}
    K. A. Bronnikov and V. N. Melnikov.
    Conformal frames and D-dimensional gravity,
    gr-qc/0310112;
    in: {\it Proc. 18th Course of the School on Cosmology and
    Gravitation: The Gravitational Constant. Generalized Gravitational
    Theories and Experiments\/} (30 April--10 May 2003, Erice),
    Ed. G.T. Gillies, V.N. Melnikov and V. de Sabbata,
    (Kluwer, Dordrecht/Boston/London, 2004) pp. 39--64.

\bibitem{infl}
    K. A. Bronnikov, S. G. Rubin, and I. V. Svadkovsky,
    Multidimensional world, inflation and modern acceleration
    Phys. Rev. D {\bf 81}, 084010 (2010).

\bibitem{Bolokhov:2010mz}
    S. V. Bolokhov, K. A. Bronnikov, and S. G. Rubin,
    Extra dimensions as a source of the electroweak model,
    Phys. Rev. D {\bf 84}, 044015 (2011).

\bibitem{Rubin:2011jm}
    S. G. Rubin and A. S. Zinger,
    The Universe formation by a space reduction cascade with random
    initial parameters,
    Gen.\ Rel.\ Grav.\ {\bf 44}, 2283 (2012); ArXiv: 1101.1274.

\bibitem{BRook}
    K. A. Bronnikov and S. G. Rubin,
    Black Holes, Cosmology and Extra Dimensions
    (World Scientific, Singapore, 2012).

\end{thebibliography}
\end{document}